# Irreversibility and hysteresis in redox molecular conduction junctions.

Agostino Migliore[*,†,‡] and Abraham Nitzan[*,†]

[†] *School of Chemistry, Tel Aviv University, Tel Aviv 69978, Israel. Phone: +972-3-6407634. Fax: +972-3-6409293.*

[‡] *Present address: Department of Chemistry, Duke University, Durham, NC 27708, USA. Phone: +1-919-6601633.*

[*] E-mails: migliore@post.tau.ac.il and nitzan@post.tau.ac.il

**CORRESPONDING AUTHOR:** Agostino Migliore. School of Chemistry, Tel Aviv University, Tel Aviv 69978 Israel. Current address: Department of Chemistry, Duke University, Durham, NC 27708, USA. Phone: +1-919-6601633. E-mails: migliore@post.tau.ac.il, agostino.migliore@duke.edu.

**ABSTRACT**

In this work we present and discuss theoretical models of redox molecular junctions that account for recent observations of nonlinear charge transport phenomena, such as hysteresis and hysteretic negative differential resistance (NDR). A defining feature in such models is the involvement of at least two conduction channels - a slow channel that determines transitions between charge states of the bridge and a fast channel that dominates its conduction. Using Marcus' theory of heterogeneous electron transfer (ET) at metal-molecule interfaces we identify and describe different regimes of nonlinear conduction through redox molecular bridges, where the transferring charge can be highly localized around the redox moiety. This localization and its stabilization by polarization of the surrounding medium and/or conformational changes can lead to decoupling of the current response dynamics from the timescale of the voltage sweep (that is, the current does not adiabatically follow the voltage), hence to the appearance of memory (thermodynamic irreversibility) in this response that is manifested by hysteresis in current-voltage cycles. In standard voltammetry such irreversibility leads to relative shift of the current peaks along the forward and backward voltage sweeps. The common origin of these behaviors is pointed out and expressions of the threshold voltage sweep rates are provided. In addition, the theory is extended (a) to analyze the different ways by which such phenomena are manifested in single sweep cycles and in ensemble averages of such cycles, and (b) to examine quantum effects in the fast transport channel.

## INTRODUCTION

Redox molecular junctions, that is junctions whose operation involves two or more oxidation states of the molecular bridge, have attracted great interest because of their ability to manifest nonlinear effects in the current-voltage response[1, 2, 3, 4, 5, 6, 7, 8, 9, 10, 11, 12] that are relevant to nanoelectronics, and to provide control mechanisms based on the connection between the charging state of the molecule and its conduction properties.[1, 2, 3, 5, 9, 11]

In a redox molecular conduction junction, the localization of the transferring charge around the redox center and its stabilization by suitable polarization of the nuclear environment can lead to weak coupling strengths to the contacts and, as a consequence, to switching between different molecular charging states by means of sequential ET processes.[13] As noted in ref 13, the existence of two (or more) locally stable charge states is not sufficient to characterize a molecular junction as redox type. Switching between them by repeated oxidation-reduction processes simply leads to current that depends on this switching rate. A prerequisite for redox junction behavior, often manifested by the appearance of negative differential resistance (NDR), hysteresis and hysteretic NDR, is the presence of a second transport channel whose conduction is large enough to determine the observed current on the one hand, and is appreciably affected by changes in the redox state of the molecule (caused by relatively slow electron exchange through the first channel) on the other. Such a mechanism characterizes recent single electron counting measurements in quantum point contacts[14, 15, 16] and has also been proposed[17] as the physical basis of NDR in spin-blockaded transport through weakly coupled-double quantum dots. While NDR and its dependence on the temperature and the nuclear reorganization after ET were the focus of the work in ref 13, the present work also considers the occurrence of hysteresis and hysteretic NDR in weakly-coupled redox junctions.

The paper is organized as follows. In next section we analyze the common underlying mechanism of irreversible effects that appear in standard voltammetry employed at single metal-molecule interfaces and hysteresis in the current-voltage response of metal-molecule-metal junctions. This analysis is then extended to redox molecular junctions characterized by two interacting, fast and slow, charge-transport channels, described by three or four molecular states models. Charge transfer kinetics in the slow channel can be safely described by sequential Marcus rate processes.[18, 19, 20] Charge transfer through the fast channel that dominates the junction current is described either using Marcus rates or as resonant tunneling according to the Landauer-Büttiker formalism.[21, 22]

## RESULTS AND DISCUSSION

### Irreversible voltammetry and hysteretic conduction in a two-state model.

In what follows we refer as *irreversible* current-voltage response the evolution of a junction that does not reverse itself when the voltage sweep is reversed. Such irreversible evolution occurs when the intrinsic charge transfer timescale (measured, *e.g.*, by $\rho^{-1}$, eq 3b below) is slow relative to the voltage sweep rate, so that the current cannot adiabatically follow the instantaneous voltage. Obviously, irreversibility in a solvated molecular junction (a double molecule-metal interface) and in cyclic voltammetry under diffusionless conditions[23] must have a similar underlying mechanism, still such studies have progressed separately so far. Several comparative observations such as (i) the behavior of single molecule conductance against the need for a molecular layer to obtain appreciable current from a volammogram[24, 25] and (ii) the appearance of irreversibility in voltammetry involving diffusionless molecules at sweep rates lower than those required for observable hysteresis in redox junctions, can be explained by addressing them together. One aim of the following analysis is to relate and explain such phenomena, by affording a common language for their description.

We start by considering the simplest molecular model: a two-state system, an oxidized molecular form $A$ and a reduced form $B$, where transitions between them take place by simple rate processes. The transition rates $A \to B$ and $B \to A$ (electron injection into and removal from the molecule, respectively) are denoted by $R_{AB} \equiv R_{A \to B}$ and $R_{BA} \equiv R_{B \to A}$, respectively. We denote by $P_A$ and $P \equiv P_B$ the probabilities to find the molecule in state $A$ and $B$, respectively, and by $P_{A,eq}$ and $P_{eq}$ their equilibrium values. Obviously, $P_A + P_B = 1$ and $P_{A,eq} R_{AB} = P_{eq} R_{BA}$ (detailed balance). Under a time dependent voltage $V(t)$ these probabilities can be written as[26]

$$P_A(V,t) = P_{A,eq}(V) - Q(V,t) = \frac{R_{BA}(V)}{R_{AB}(V) + R_{BA}(V)} - Q(V,t) \tag{1a}$$

and

$$P(V,t) = P_{eq}(V) + Q(V,t) = \frac{R_{AB}(V)}{R_{AB}(V) + R_{BA}(V)} + Q(V,t), \tag{1b}$$

where the departure $Q$ of $P$ from $P_{eq}$ depends explicitly on the time $t$. All the memory effects in the response of the system to the external bias $V$ can be encapsulated in the function $Q$, which is obtained as follows: the master equation

$$\frac{dP(V(t),t)}{dt} = u\frac{dP_{eq}}{dV} + \frac{dQ}{dt} = (1-P)R_{AB}(V) - PR_{BA}(V)$$
$$= R_{AB}(P_{A,eq} - Q) - R_{BA}(P_{B,eq} + Q) = -(R_{AB} + R_{BA})Q \tag{2}$$

where $u = dV/dt$ is the rate of the voltage sweep, is rewritten as

$$\frac{dQ}{dt} = -\rho\left[Q + \frac{u}{\rho}\frac{dP_{eq}}{dV}\right], \tag{3a}$$

where

$$\rho = R_{AB} + R_{BA} \tag{3b}$$

is the effective rate that characterizes the system relaxation after changing the external voltage. For constant $u$ or over a time interval in which $u$ does not change appreciably this leads to

$$Q(t) = Q(t_0)e^{-\rho(t-t_0)} - u\int_{t_0}^{t} dt' e^{-\rho(t-t')} \left(\frac{dP_{eq}}{dV}\right)_{t'} \qquad (t \geq t_0) \tag{4}$$

The transient associated with $Q(t_0)$ can be disregarded at long time. If $u$ is small enough so that $dP_{eq}/dV$ remains essentially constant during a time interval comparable with $1/\rho$, eq 4 results in

$$Q(V,t) \cong -\frac{u}{\rho}\frac{dP_{eq}}{dV} \tag{5}$$

Eq 5 describes a steady-state value of $Q$: the difference $Q = P - P_{eq}$ remains very close to zero while $V$ is slowly changed.[27] Then, at a single molecule-metal interface under reversible conditions the current $I$ between the metal and the molecule is proportional to $u$ and is given by

$$J \equiv \frac{I}{e} = u\frac{dP_{eq}}{dV} = -\rho Q \equiv -\rho(P - P_{eq}) \tag{6}$$

where $e$ is the magnitude of the electron charge.

Irreversibility manifests itself in accumulation of $Q$ during part of a voltage sweep and inversion in the sign of $Q$ in the backward sweep, with consequent hysteresis over a cycle. The general requirement for reversible behavior at any $V$ is obtained from eqs 1 and 6 as

$$\frac{u}{\rho P_{eq}}\frac{dP_{eq}}{dV} = \frac{-Q}{P_{eq}} \ll 1 \tag{7}$$

and can be extended to models of single or double metal-molecule interfaces that include more than two system states (see next section).

For a molecule stably adsorbed on a single metal electrode that can be characterized as a semi-junction, eq 2 or 6 can be used to describe on a "per molecule" basis[28, 29, 30] the current at a molecule-electrode interface, as it appears in typical linear scan (cyclic) voltammograms of diffusionless redox systems. In such a system, the applied overpotential $V$ operates as a gate voltage, effectively changing the position of the molecular level relative to the metal Fermi energy. The surface concentration of the electroactive species in the reduced (oxidized) state is replaced by the occupation probability $P_B = P$ ($P_A = 1 - P$) of the molecular redox site, and its time derivative yields the charge flow originating from the change in the oxidation state of the molecular system.[31] A fast enough voltage sweep leads to the hysteretic behavior of P shown in Figure 1a, which was obtained by implementing eq 2 in a finite difference simulation and describing the interfacial ET according to the Gurney[32]-Marcus model, as in ref 29, but with the ET rates in the analytical form reported in the Appendix. We assign $V$ as positive when the electrostatic potential on the molecule is higher than that in the metal,[13] so that electrons flow from the metal, making it identical with the negative of the traditional definition of the overpotential

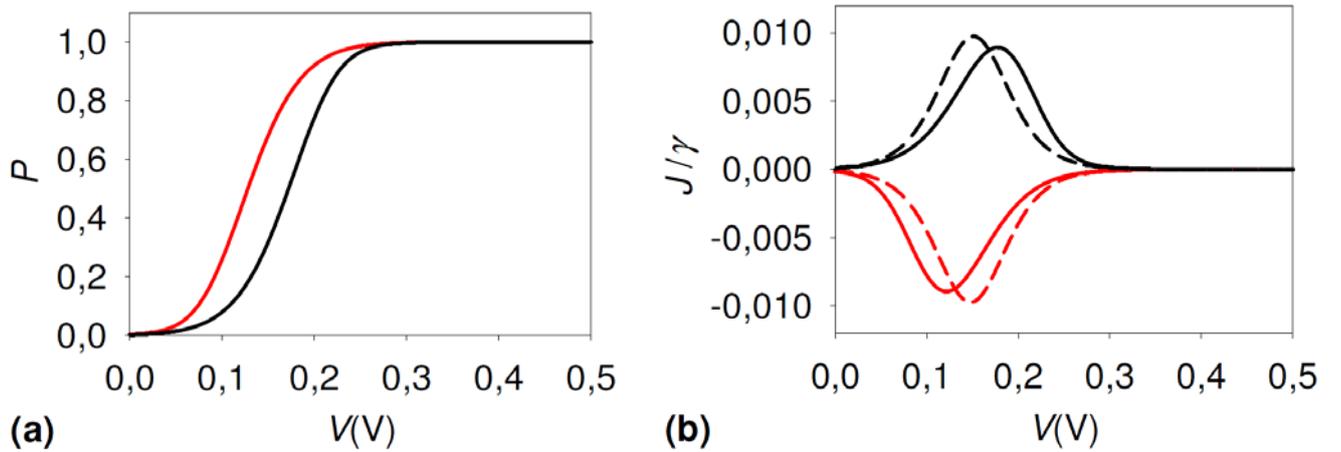

**Figure 1.** (a) The occupation probability $P$ of the molecular site plotted against the interfacial voltage $V$ over a cycle, with a maximum voltage $V_{max} = 0.5\,\text{V}$. The forward and backward sweeps are in black and red, respectively. The system is modeled by the parameters $T = 298\,\text{K}$, $E_{AB} - \mu = 0.15\,\text{eV}$, and $\lambda = 0.25\,\text{eV}$. The initial condition is $P(0) = P_{eq}(0)$. The scan rate is given by $u/\gamma = 10^{-3}\,\text{V}$. For example, $u = 10\,\text{V/s}$ for $\gamma = 10^4\,\text{s}^{-1}$ or $u = 100\,\text{V/s}$ for $\gamma = 10^5\,\text{s}^{-1}$ (b) The dimensionless current at the molecule-metal interface, $J/\gamma = dP/(\gamma\,dt)$, plotted against $V$ using eq 2. The solid lines are obtained with the same parameters as in panel a, in particular $\lambda = 0.25\,\text{eV}$. The dashed lines correspond to the same parameters, except that $\lambda = 0$.

Figure 1 relates the splitting of the peak potentials to the hysteresis in $P$. Such a connection, which is not explicitly considered in theoretical analyses of electrochemical redox reactions and voltammetry, is used here to link the peak splitting observed in single irreversible cyclic voltammograms (see Figure 1b, where the average response of the single adsorbed molecule over many sweeps can be compared to the voltammogram for a molecular layer) and the hysteretic $I$-$V$ characteristics of redox junctions.

As investigated both theoretically[33, 34] and experimentally,[35] the time derivative of $P_{eq}$ in eq 6 departs from the ideal behavior (characterized by a peak of size $eu/4k_BT$ that occurs at equal potentials in the upward and downward scans) when the voltage scan rate is comparable with the ET rate[20] as quantified by the dimensionless kinetic parameter[28]

$$m = k^0 \, k_B T / eu \qquad (8)$$

where $k^0 = R_{AB}(V=0) = R_{BA}(V=0)$ with $E_{AB} = \mu$. For the purpose of this work it important to describe this distortion (and the associated hysteresis in the redox state of the molecule) in terms of a kinetic parameter that lends itself to generalization and use within the context of redox molecular junctions. In particular, the explicit dependence on the reorganization energy $\lambda$ and on the voltage $V$ needs to be expressed. Our aim is to provide a criterion for the first appearance of hysteresis. To this aim, we consider that $P$ experiences the largest rate of change, hence the first appearance of hysteresis (see Figure 1a), about the voltage $V_0 = (E_{AB} - \mu)/e$ at which $R_{AB} = R_{BA}$ (see eq 34 and Figure 2). At this voltage, the effective rate $\rho$ can be written as (see Appendix and ref 36)[37]

$$\rho(V_0; \gamma, \lambda, T) = \begin{cases} \gamma & (\lambda = 0) \\ \gamma \sqrt{\dfrac{\pi k_B T}{\lambda}} \exp\left(-\dfrac{\lambda}{4 k_B T}\right) & (\lambda \gg k_B T) \end{cases} \qquad (9)$$

Near $V_0$, $\rho$ decreases with the reorganization energy (for example, see Figure 5 in ref 38), so that eq 6 can become invalid at any feasible scan rate and irreversible behavior is observed. In contrast, the condition for reversibility can be much more easily satisfied at usual sweep rates (up to $\sim 100\,\text{V/s}$) for $\lambda = 0$, namely, for $\rho = \gamma$. This is exemplified in Figure 2: for $\lambda = 0.25\,\text{eV}$, the evolution of $P$ over a cycle of the applied voltage is characterized by hysteresis with a corresponding splitting of the peak voltages in Figure 1b, whereas no hysteresis occurs for $\lambda = 0$. This can be quantified: inserting the expression of the maximum current (in reversible regime), $eu/4k_BT$, and eq 9 into eq 6, and imposing the condition $Q \ll 1/2$ for reversibility, we arrive at the limiting sweep rate

$$u_l(V_0;\gamma,\lambda,T) = 2\frac{k_BT}{e}\rho(V_0;\gamma,\lambda,T), \qquad (10)$$

such that reversible behavior is obtained if $u \ll u_l$, while hysteresis in the evolution of $P$ over a voltage cycle, hence distortion of the voltammogram, takes place if $u \gtrsim u_l$.

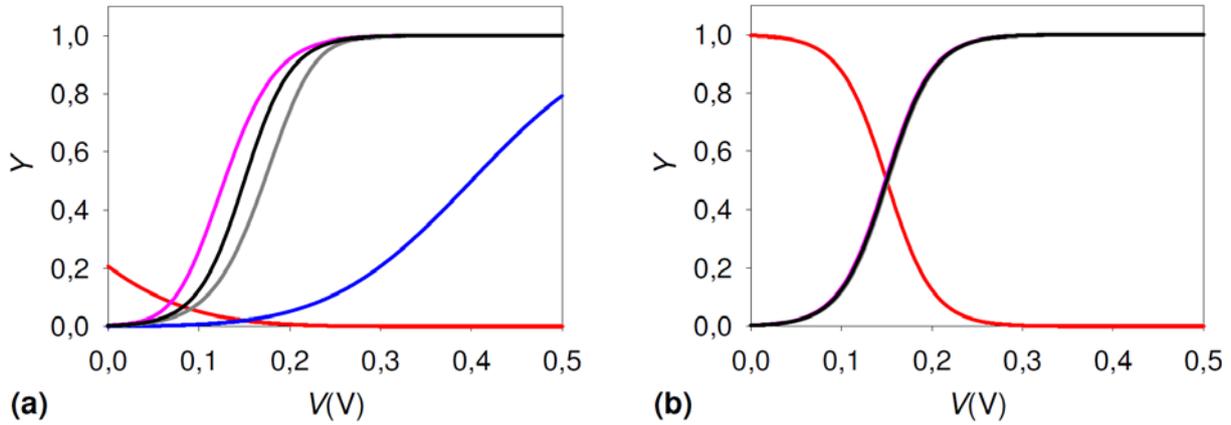

**Figure 2.** (a) $Y = R_{BA}/\gamma$ (red), $R_{AB}/\gamma$ (blue) and $P_{eq}$ (black) plotted against $V$, and $P$ along the forward (gray) and backward (pink) voltage sweeps, by using the same model parameters as in Figure 1a. (b) The same quantities as in panel a are shown for the case $\lambda = 0$ (as in Figure 1b).

While eq 10 yields a delimiter between the sweep rate ranges with reversible and irreversible behavior based on the first appearance of hysteresis at $V = V_0$, its extension to all voltages is obtained by application of eq 7 and use of the rate expressions derived in the Appendix, as:

$$u_l(V;\gamma,\lambda,T) = 2\gamma\frac{k_BT}{e}\tilde{\rho}(V;\lambda,T), \qquad (11a)$$

with

$$\tilde{\rho}(V;\lambda,T) = \begin{cases} 1 & (\lambda = 0) \\ \dfrac{S(\lambda,T,\alpha)}{8}\left[1+\exp\left(-\dfrac{\alpha}{k_B T}\right)\right]^2 \\ \times \exp\left[-\dfrac{(\lambda-\alpha+2\sqrt{\alpha\lambda})(\lambda-\alpha-2\sqrt{\alpha\lambda})}{4\lambda k_B T}\right] & (\lambda \neq 0). \end{cases} \quad (11b)$$

where $\alpha$ is defined by eq 34b. Eqs 10 and 11 provide generalizations of the dimensionless rate constant $m$ given by eq 8 and employed in standard voltammetry studies. Eq 11 defines the lower boundary of the irreversible region in the $V$-$u$ plane as the locus $m(V) \equiv 2\gamma\tilde{\rho}(V)k_B T/eu(V) = 1$, while eq 10 yields an approximation to the maximum of such a curve. These equations can be used in future analyses for full theoretical characterization of the intermediate behaviors between the reversible ($m \to \infty$) and totally irreversible ($m \to 0$) limits (set in the fundamental work by Laviron[28, 34, 39] by using Butler-Volmer equations) with use of Marcus ET rates and thus consideration of reorganization energy and temperature effects. In particular, according to above interpretation of eqs 10 and 11 in the $V$-$u$ plane, the condition

$$u_l(V_0;\gamma,\lambda,T) \lesssim u \ll 2\gamma\frac{k_B T}{e} \quad (12)$$

defines a regime of sweep rates where irreversibility is seen exclusively in the presence of suitably large reorganization energy. For example, for $\lambda = 0.25\,\text{eV}$ eq 10 gives $m = u_l/u \cong 0.4$, and in fact irreversible response is found in Figures 1-2, while for $\lambda = 0$ it is $u_l/u \cong 20$ and no irreversibility occurs. Eqs 10-12 can be applied, e.g., to the $\gamma$ range from $\sim 10\,\text{s}^{-1}$ to $\sim 10^6\,\text{s}^{-1}$ deduced in the Supporting Information from experimental data, and can be used to explore and predict the effects on irreversible behaviors of using solvents with diverse polarization properties, hence different resultant reorganization energy, in distinct experiments.

Further discussion of novelty and significance of eqs 9-12 is afforded in the Supporting Information.

Next consider the double metal-molecule interface of a redox molecular conduction junction. Here $P_{eq}$ in eqs 1a-b is replaced by $P_{ss}$ the (non-equilibrium) steady state probability that the system is in state $B$. Eqs 1-5 also apply to the two-state model of such a junction, where both the left (denoted by $L$) and right ($R$) contacts, characterized by coupling the strengths $\gamma^L$ and $\gamma^R$, respectively, contribute to the transitions $A \to B$ and $B \to A$, so that the respective ET rate constants are given by $R_{AB} = R_{AB}^L + R_{AB}^R$ and $R_{BA} = R_{BA}^L + R_{BA}^R$. Considering, for simplicity, a symmetric junction ($\gamma^L = \gamma^R = \gamma$) and symmetric bias drop at the electrodes, the ET rates are still given by eq 34 except that $\alpha$ is replaced by $\alpha_K \equiv \mu - E_{AB} - e\phi_K$ ($K = L, R$), where $\phi_R = V/2 = -\phi_L$.[13] The instantaneous $L$- and $R$-terminal currents are given by

$$J_L \equiv \frac{I_L}{e} = (1-P)R_{AB}^L - P R_{BA}^L = (P_{A,ss} - Q)R_{AB}^L - (P_{ss} + Q)R_{BA}^L = J_{AB} - (R_{AB}^L + R_{BA}^L)Q, \quad (13a)$$

$$J_R \equiv \frac{I_R}{e} = P R_{BA}^R - (1-P)R_{AB}^R = J_{AB} + (R_{AB}^R + R_{BA}^R)Q, \quad (13b)$$

where

$$J_{AB} = \frac{R_{AB}^L R_{BA}^R - R_{AB}^R R_{BA}^L}{R_{AB} + R_{BA}} \quad (13c)$$

is the steady-state current. Deviation from steady-state can be expressed by the difference between the left and right terminal currents:[40]

$$J_L - J_R = -(R_{AB} + R_{BA})Q = \frac{dP}{dt}. \quad (14)$$

While in the single interface case discussed above $dP/dt$ is the interfacial current, here $dP/dt$ is a "leakage" current that vanishes under steady-state conditions. The reversible/irreversible behavior of the junction, as expressed by hysteresis in the current over a bias cycle, can be described as before: irreversibility sets in when the bias sweep rate $u = V/t$ is larger than the current relaxation rate determined by $\rho = R_{AB} + R_{BA}$, and it can be conveniently described in terms of $Q = P - P_{ss}$. Eqs 3 and 5 remain valid also in the present case, because they refer to a generic two-state system, and, together with eqs 13a-b, provide the following criterion of reversibility (analogous to eq 7):

$$|J_L - J_R| \ll P_{ss}\rho = R_{AB} \qquad (15)$$

where

$$J_L - J_R = -\rho Q = u\frac{dP_{ss}}{dV} \qquad (16)$$

Since $R_{AB}$ determines the order of magnitude of $J_L$, and $J_R$, it follows also that $|J_L - J_R| \ll J_L, J_R$ in this limit. Eqs 15 and 16 yield the following condition on the sweep rate for the attainment of reversible current-voltage responses:

$$u \ll u_0 \equiv \frac{\rho P_{ss}}{|dP_{ss}/dV|} = \frac{\rho^2}{\left|\frac{R_{BA}}{R_{AB}}\frac{dR_{AB}}{dV} - \frac{dR_{BA}}{dV}\right|} \qquad (17)$$

As detailed in the Supporting Information, $u_0$ decreases with $\lambda$ at any voltage. This may suggest the hysteresis in the current-voltage response of a redox molecular junction is easier to detect with feasible scan rates when the reorganization energy involved in the electron localization on the redox center is larger. However, Figure 3 shows that this conclusion is too simplistic because, in contrast with the trend in the irreversible behavior of a molecule adsorbed on a single electrode, increasing $\lambda$ not only makes $u_0$ smaller but also makes the hysteresis cycle narrower.[41]

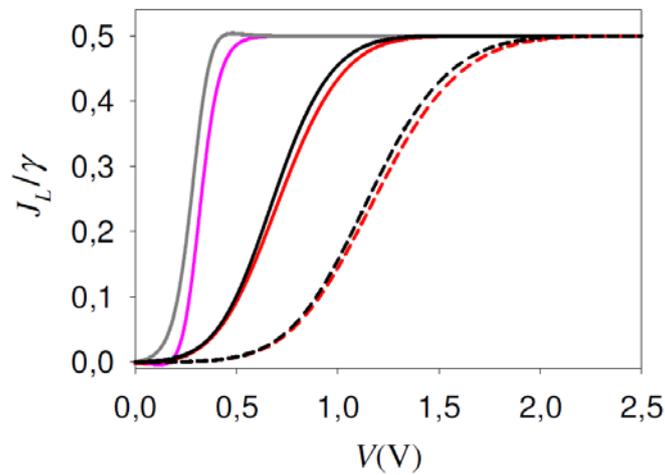

**Figure 3.** $J_L/\gamma$ plotted against $V$, for $T = 298\,\text{K}$, $E_{AB} - \mu = 0.15\,\text{eV}$, and a voltage sweep rate that for $\gamma = 300 \text{s}^{-1}$ is $u = 20 \text{V/s}$. It is $\lambda = 0$ (forward and backward sweeps in grey and pink, respectively) $\lambda = 0.25 \text{ eV}$ (solid black and red lines), and $\lambda = 0.5 \text{ eV}$ (black and red dashes). A small transient NDR is seen for $\lambda = 0$ due to fast charge accumulation in the molecule, as given by $dP/dt$.[41]

To conclude this section, we consider again the timescale issue. As already stated, irreversibility and hysteresis occur when the current cannot adiabatically follow the voltage change, which requires that the characteristic charge transfer rates are slower than the voltage scan rate. On the other hand, in a single-molecule junction easily observable currents (*i.e.*, currents of the order of 1 nA) require the transit of $\sim 10^{10}$ electrons/s. Thus, the condition for detectable current is clearly incompatible with the condition for hysteresis with experimentally feasible scan rates. The model considered so far, *i.e.*, a two-state molecular junction, cannot account for such experimental observations.[42] A four-state model (that becomes a three-state model over voltage ranges where double occupation of the bridge is not allowed) able to justify the occurrence of significant hysteresis under less restrictive conditions is presented in the next section.

**Redox molecular junctions.**

The occurrence of hysteresis, NDR, and hysteretic NDR at sweep rates commonly used in experiments can be rationalized even in single molecule junctions, provided that charge transport through the molecule takes place via at least two channels with different characteristics: one (strongly coupled or "fast") channel dominates the observed current, while the other (weakly coupled, "slow") channel determines that charging state of the bridge. In ref 13 we have argued that the existence of two such channels is the hallmark of so called redox molecular junctions.

For definiteness, we consider the neutral molecule (state *A*) and two single-electron orbitals, *b* and *c*, that can become occupied when the molecule acquires excess electron(s). We assume that orbital *c* is strongly localized on the molecule (as would be the case for an orbital localized near a redox center), therefore weakly coupled to at least one of the electrodes, while orbital *b* is more delocalized, so more strongly coupled to both electrodes. In the ensuing kinetics orbital *b* will provide a relatively fast channel that determines the magnitude of the observed current, while population and depopulation of orbital *c* takes place on a slow time scale associated with the observed hysteresis and NDR. Figure 4 depicts this model in the molecular state space. Molecular states *B* and *C* correspond to the molecule with an excess electron in orbital *b* and *c*, respectively, while the state where both orbitals are occupied is denoted by *D*.[43] A similar model, excluding population of state *D*, has been used by Muralidharan and Datta,[17] who proposed a mechanism for NDR in the Coulomb blockade limit, and in works by Flensberg et al.,[44, 45] where it is shown that the blocking state causing NDR can result by breaking of the molecular symmetries due to image charge interaction. Transport models that comprise interacting fast and slow channels have been also studied recently in the context of electron counting measurements, where the current through a point contact is used to monitor the electron occupation in a neighboring weakly transmitting junction.[14, 15, 46] When applied to redox molecular junctions, such models have to take into account strong electron-phonon coupling and the dynamics of nuclear reorganization, which is done here by inserting Marcus-type interface ET rates in simple rate (master) equations for both the slow and fast transport channel.

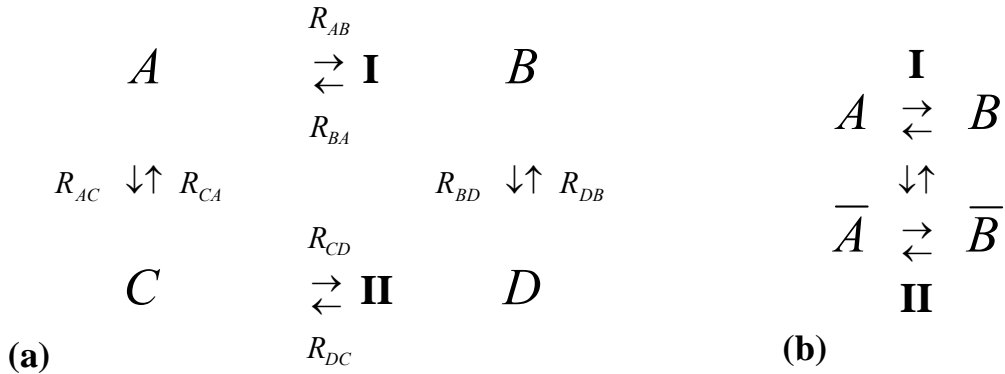

**Figure 4.** (a) The four-state model described in the text. The transition rate for the process $W \to X$ ($W$, $X = A, B, C, D$) is denoted by $R_{WX}$. I and II denote the conduction modes of the "fast" transport channel in the two oxidation states. The vertical arrows depict the changes in the molecular oxidation state by electron transfer via the "slow" channel that often involves transient localization of excess charge in the redox group. (b) Same as (a), in a reduced picture where $\bar{A}$ and $\bar{B}$ are obtained from states *A* and *B*, respectively, by charging the redox site.

It should be noted that the applicability of such a kinetic description using Marcus ET rates is not at all obvious. It is certainly justified for the slow channel, where the timescale for molecular charging and discharging is slow relative to that of nuclear rearrangement, however it may be questionable for the charge transfer transitions associated with the fast channel. In next section, we will consider a novel conduction model where the hopping mechanism is assumed for the slow channel while the fast channel is described in the Landauer-Büttiker limit (coherent transport), as it may be appropriate depending on the metal-molecule coupling strengths. Here we continue to assume that both channels can be described with Marcus hopping kinetics. The corresponding master equation is

$$\begin{cases} \dfrac{dP_A}{dt} = -(R_{AB} + R_{AC})P_A + R_{BA}P_B + R_{CA}P_C \\ \dfrac{dP_B}{dt} = -(R_{BA} + R_{BD})P_B + R_{AB}P_A + R_{DB}P_D \\ \dfrac{dP_C}{dt} = -(R_{CA} + R_{CD})P_C + R_{AC}P_A + R_{DC}P_D \\ \dfrac{dP_D}{dt} = -(R_{DB} + R_{DC})P_D + R_{BD}P_B + R_{CD}P_C \end{cases} \quad (19)$$

where $P_X$ is the probability of the molecular state $X$ ($X = A, B, C$ and $D$) and $P_A + P_B + P_C + P_D = 1$. The total probability of charge localization in the redox site will be denoted by $P$, namely, $P \equiv P_C + P_D$.. The transport process given by eq 19 can be described in different ways. First, in a single-electron picture, orbitals $b$ and $c$ describe two different distributions of a transferring electron on the molecule, which correspond to the two transport channels discussed above. These channels will be denoted by 1 and 2, respectively. Alternatively, because of the vastly different timescales associated with these channels, and because channel 2 contributes negligibly to conduction, we can consider two conduction modes of channel 1 that correspond to the different occupation states of orbital $c$. In the molecular state language, these conduction modes, denoted by I and II in Figure 4, correspond to the $A \leftrightarrow B$ and $C \leftrightarrow D$ processes, respectively. We will sometime simplify the notation further, denoting these processes by $A \leftrightarrow B$ and $\overline{A} \leftrightarrow \overline{B}$, where $\overline{A}$ and $\overline{B}$ represent the molecular states of the "charged" (in the sense that $c$ is occupied) molecule in which orbital $b$ is empty or occupied, respectively, as seen in Figure 4b. Note that in most calculations reported below we also take into account the small contribution to the current from channel 2.

The ET rates in eq 19 are given by expressions similar to eq 34, except that the state energies and molecule-electrode coupling parameters are adjusted to take into account electron-electron interaction as expressed in the properties of the molecular states *B*, *C* and *D*. Specifically, the state energies satisfy $E_D - E_A = (E_B - E_A) + (E_C - E_A) + \chi$, where $\chi$ is the energy of interaction between the two excess charges in state *D*. In general, $\chi \neq 0$ and the $C \leftrightarrow D$ and $B \leftrightarrow D$ transition rates depend on the energy differences $E_{CD} \equiv E_D - E_C = E_{AB} + \chi$ and $E_{BD} = E_{AC} + \chi$, respectively. Furthermore, we consider the possible effects of charging one channel on the properties of the other channel. Denoting by $\lambda_i$ and $\gamma_i^K$ (*i* = 1, 2) the reorganization energy[47] and the coupling strength (expressed by the corresponding electron loss rate) to the *K* (= *L*, *R*) contact in channel *i* in the case where $\chi = 0$, we neglect the effect of occupying the (relatively delocalized) *b* orbital on the localized *c* wave function, hence on the parameters $\gamma_2^L$, $\gamma_2^R$ and $\lambda_2$ associated with this orbital, as suggested by recent studies based on the Density-Functional theory.[48, 49] In contrast, charge localization in *c* causes significantly inhomogeneous spatial changes in the effective potential seen by the other transferring charge, with non-negligible effects on orbital *b*. In particular, a change in the wave function tails on the two electrodes may lead to significant changes in the metal-molecule electronic couplings. This is modeled by assigning the coupling strengths $\bar{\gamma}_1^K = \kappa \gamma_1^K$ (with $\kappa$ a constant and *K* = *L*, *R*) to channel 1 in the conduction mode II. On the other hand, we disregard a possible effect of charging orbital *c* on $\lambda_1$ (see the inclusion of this effect in the Supporting Information).

To summarize, the essential features of the above four-state model for a redox molecular junction are:

(a) Such junctions are characterized by two conduction channels: a fast channel, 1, and a slow channel, 2. The charging transitions in the latter are dominated by electron localization at a molecular redox center.

(b) The transitions between the charged and uncharged states of the redox group (slow channel) contribute negligibly to the junction current but can affect significantly the conductance via channel 1.

(c) The timescale separation between molecular charging (transitions in channel 2) and conduction through channel 1 results in transitions between junction states characterized by different steady-states currents.

The timescale separation between the two channels is the essential attribute of the redox junction property, which is amplified by the solvent reorganization about the redox site. It has the important consequence that the decoupling of the timescale associated with charging/discharging of the molecular redox site and that of the voltage scan occurs at scan rates far slower than the charge transfer through the fast channel that dominates the junction current. Therefore, in contrast to the two-state case, irreversibility and memory effects in the population kinetics of the redox center will be expressed visibly, sometimes prominently, in the observed conduction. In particular, hysteresis and hysteretic NDR will be dominated by the slow channel, and telegraphic noise associated with transitions in this channel is expected in some voltage range. Note that because of the large reorganization energy associated with channel 2, its effect on the molecular conduction begins at voltages higher than the threshold for significant current through channel 1.[13]

Next consider the junction transport properties as described by eqs 19. The solution of these equations is greatly simplified by exploiting the timescale separation between the two channels. Assuming (see point (c) above) that channel 1 is at steady-state, so that

$$P_A R_{AB} = P_B R_{BA} \tag{20a}$$

$$P_C R_{CD} = P_D R_{DC} \tag{20b}$$

eqs 19 reduces to

$$\begin{cases} \dfrac{dP_A}{dt} = -\dfrac{dP_C}{dt} = -R_{AC} P_A + R_{CA} P_C \\ \dfrac{dP_B}{dt} = -\dfrac{dP_D}{dt} = -R_{BD} P_B + R_{DB} P_D \end{cases} \tag{21}$$

The left terminal current, normalized to $e$, is given by

$$J_L(V;\chi) = P_A R_{AB}^L + P_C R_{CD}^L + P_A R_{AC}^L + P_B R_{BD}^L - \left( P_B R_{BA}^L + P_D R_{DC}^L + P_C R_{CA}^L + P_D R_{DB}^L \right) \tag{22}$$

Using eqs 20, 22, 34 and the relation $P \equiv P_C + P_D$ allows us to write this current as a function of the voltage $V$ and the interaction parameter $\chi$ in the form (see Supporting Information)

$$J_L(V;\chi) = J^P(V;\chi) + \Lambda_L(V;\chi), \qquad (23a)$$

where

$$J^P(V;\chi) = [1 - P(V;\chi)]J_{AB}(V) + P(V;\chi)r(V;\chi)J_{AB}(V) \qquad (23b)$$

with

$$r(V;\chi) \equiv \frac{J_{\overline{AB}}}{J_{AB}} = \kappa \left\{ \left[1 + \exp\left(-\frac{\alpha_L}{k_B T}\right)\right] R_{AB}^L + \left[1 + \exp\left(\frac{\alpha_R}{k_B T}\right)\right] R_{BA}^R \right\}$$

$$\left/ \left\{ \frac{S(\lambda_1, T, \alpha_R)}{S(\lambda_1, T, \alpha_R - \chi)} \exp\left[\frac{\chi\left(\frac{\chi}{2} - \alpha_R - \lambda_1\right)}{2\lambda_1 k_B T}\right] \left[1 + \exp\left(-\frac{\alpha_L - \chi}{k_B T}\right)\right] R_{AB}^L \right. \right. \qquad (23c)$$

$$\left. + \frac{S(\lambda_1, T, \alpha_L)}{S(\lambda_1, T, \alpha_L - \chi)} \exp\left[\frac{\chi\left(\frac{\chi}{2} - \alpha_L + \lambda_1\right)}{2\lambda_1 k_B T}\right] \left[1 + \exp\left(\frac{\alpha_R - \chi}{k_B T}\right)\right] R_{BA}^R \right\}$$

is the contribution to the current by channel 1. In eq 23c, $J_{\overline{AB}}$ (or, equivalently, $J_{CD}$; see Figure 4) denotes the steady-state current carried by channel 1 through the reduced molecule (a molecule with an excess electron localized at the redox center). It is given, in analogy to eq 13c, by

$$J_{\overline{AB}} = \frac{R_{\overline{AB}}^L R_{\overline{BA}}^R - R_{\overline{AB}}^R R_{\overline{BA}}^L}{R_{\overline{AB}} + R_{\overline{BA}}} \qquad (23d)$$

where the ET rates are given by the analogue of eq 34 for the conduction mode II of channel 1 (*i.e.*, using $E_{CD}$ rather than $E_{AB}$ in eq 34 of the Appendix). The second term in the right side of eq 23a is the small contribution to the current by channel 2, given by

$$\Lambda_L(V;\chi) = (1-P)\frac{R_{AB}R_{BD}^L + R_{BA}R_{AC}^L}{R_{AB} + R_{BA}} - P\frac{R_{CD}R_{DB}^L + R_{DC}R_{CA}^L}{R_{CD} + R_{DC}}$$

$$= (1-P)\frac{\eta_{AC}^L R_{AB} + R_{BA}}{R_{AB} + R_{BA}} R_{AC}^L - P\frac{\eta_{AC}^L(\eta_{AB}^L R_{AB}^L + \eta_{AB}^R R_{AB}^R) + \eta_{AB}^L R_{BA}^L + \eta_{AB}^R R_{BA}^R}{(\eta_{AB}^L R_{AB}^L + \eta_{AB}^R R_{AB}^R)\exp\left(-\frac{\chi}{k_B T}\right) + \eta_{AB}^L R_{BA}^L + \eta_{AB}^R R_{BA}^R} R_{CA}^L \quad (23e)$$

with

$$\eta_{AB}^K(V;\chi) \equiv \frac{R_{CD}^K(V;\chi)}{R_{AB}^K(V)} = \kappa \frac{S(\lambda_1, T, \alpha_K - \chi)}{S(\lambda_1, T, \alpha_K)} \exp\left[\frac{\chi(\alpha_K - \lambda_1 - \chi/2)}{2\lambda_1 k_B T}\right], \quad (K = L, R) \quad (23f)$$

and $\eta_{AC}^L$ expressed similarly to $\eta_{AB}^L$ by replacing $\kappa$, $\lambda_1$ and $E_{AB}$ with 1, $\lambda_2$ and $E_{AC}$, respectively. For $\chi = 0$, $\Lambda_L$ takes the simple form $(1-P)R_{AC}^L - PR_{CA}^L$ and under steady-state conditions it is given by eq 13c with $B$ replaced by $C$. This small contribution to the current is disregarded in Figure 4b.

Eq 23b expresses $J^P$, the dominant contribution to the current $J_L$ at the left electrode, as a weighted average of the currents carried by channel 1 in its conduction modes I and II. When the interaction between the two excess electron charges in state $D$ is neglected, the ratio $r$ of the steady-state currents $J_{\overline{AB}}$ and $J_{AB}$ is voltage independent, $r(V;0) = \kappa$, as seen from eqs 13c and 23f. An analogous expression can be written for the $R$-terminal current, $J_R$:

$$J_R(V;\chi) = J^P(V;\chi) + \Lambda_R(V;\chi) \quad (24a)$$

with

$$\Lambda_R(V,\chi) = -(1-P)\frac{\eta_{AC}^R R_{AB} + R_{BA}}{R_{AB} + R_{BA}} R_{AC}^R$$

$$+ P\frac{\eta_{AC}^R(\eta_{AB}^L R_{AB}^L + \eta_{AB}^R R_{AB}^R) + \eta_{AB}^L R_{BA}^L + \eta_{AB}^R R_{BA}^R}{(\eta_{AB}^L R_{AB}^L + \eta_{AB}^R R_{AB}^R)\exp\left(-\frac{\chi}{k_B T}\right) + \eta_{AB}^L R_{BA}^L + \eta_{AB}^R R_{BA}^R} R_{CA}^R \quad (24b)$$

The memory effects that appear in fast sweeps are embodied into eqs 23-24 through the deviation $Q$ of $P$ from its steady-state value $P_{ss}$. The evolution of $P$ is derived from eq 21 (see Supporting Information) as

$$\frac{dP}{dt} = u\frac{dP_{ss}}{dV} + \frac{dQ}{dt} = -A_2 Q \qquad (25)$$

(the dependence of $P$ on $V = ut$ and $\chi$ is not explicitly shown here), where

$$P_{ss} = \frac{A_1}{A_2} = \frac{1}{1 + \dfrac{R_{AB} + R_{BA}}{R_{CD} + R_{DC}} \dfrac{R_{CD} R_{DB} + R_{DC} R_{CA}}{R_{AB} R_{BD} + R_{BA} R_{AC}}}, \qquad (26a)$$

with

$$A_1 = \frac{R_{BA}}{R_{AB} + R_{BA}} R_{AC} + \frac{R_{AB}}{R_{AB} + R_{BA}} R_{BD} = \frac{R_{AC}}{R_{AB} + R_{BA}}\left(R_{BA} + R_{AB}\frac{\eta_{AC}^L R_{AC}^L + \eta_{AC}^R R_{AC}^R}{R_{AC}}\right) \qquad (26b)$$

$$A_2 = \frac{R_{BA}}{R_{AB} + R_{BA}} R_{AC} + \frac{R_{AB}}{R_{AB} + R_{BA}} R_{BD} + \frac{R_{DC}}{R_{CD} + R_{DC}} R_{CA} + \frac{R_{CD}}{R_{CD} + R_{DC}} R_{DB}$$

$$= A_1 + \frac{R_{CA}}{\eta_{AB}^L R_{AB}^L + \eta_{AB}^R R_{AB}^R + \left(\eta_{AB}^L R_{BA}^L + \eta_{AB}^R R_{BA}^R\right)\exp\left(\dfrac{\chi}{k_B T}\right)} \qquad (26c)$$

$$\times \left[\eta_{AB}^L R_{BA}^L + \eta_{AB}^R R_{BA}^R + \left(\eta_{AB}^L R_{AB}^L + \eta_{AB}^R R_{AB}^R\right)\frac{\eta_{AC}^L R_{CA}^L + \eta_{AC}^R R_{CA}^R}{R_{CA}}\right]\exp\left(\dfrac{\chi}{k_B T}\right)$$

and $\eta_{AC}^R$ obtained from $\eta_{AB}^R$, eq 23f, by replacing $\kappa$, $\lambda_1$ and $E_{AB}$ with 1, $\lambda_2$ and $E_{AC}$, respectively.

From eqs 23-26 it follows that if the sweep rate is significantly smaller than

$$u_0 = \frac{A_2 P_{ss}}{dP_{ss}/dV}. \qquad (27)$$

channel 2 also works under steady-state conditions, so that $P = P_{ss}$ and the *I-V* characteristics of the junction does not exhibit any hysteretic behavior. In this case, $\Lambda_L = \Lambda_R$ and consequently $J_L = J_R$. If, instead, $u$ is similar to or smaller than $u_0$ in some bias range, $Q$ is an appreciable fraction of $P_{ss}$ and irreversibility appears in $J_L$ and $J_R$ through both the main contribution $J^P$ and the residual terms $\Lambda_L$ and $\Lambda_R$. Still, since $J^P \gg \Lambda_R, \Lambda_L$, $J_L - J_R = \Lambda_L - \Lambda_R = dP/dt$ is much smaller than $J_L$ and $J_R$, so that $J_L \cong J_R \cong J^P$. It is worth noting that the kinetic parameter $m$ introduced in eq 8 is extended to the present model by replacing $k^0 k_B T/e$ with the limiting voltage sweep rate $u_0$ of eq 27. Then, $u = u_0$ amounts to the condition $m(V) = 1$ that was discussed for the two-state case.

Figures 5-9 show some characteristic behaviors resulting from Eqs 23, 25 and 26 (for specificity, only the current $J_L$ is shown). In these examples we assume equal potential drops across the two molecule-lead interfaces.[50] Figure 5 shows the occurrence of hysteresis, NDR, and hysteretic NDR for a given set (see caption) of junction parameters. As seen in Figures 5c and 5f, high enough scan rates lead to hysteresis irrespective of the value of the electron-electron interaction parameter $\chi$. On the other hand, large enough $\chi$ causes NDR irrespective of the scan rate (see Figures 5d-f). Thus, hysteretic NDR occurs for sufficiently high values of both $u$ and $\chi$.

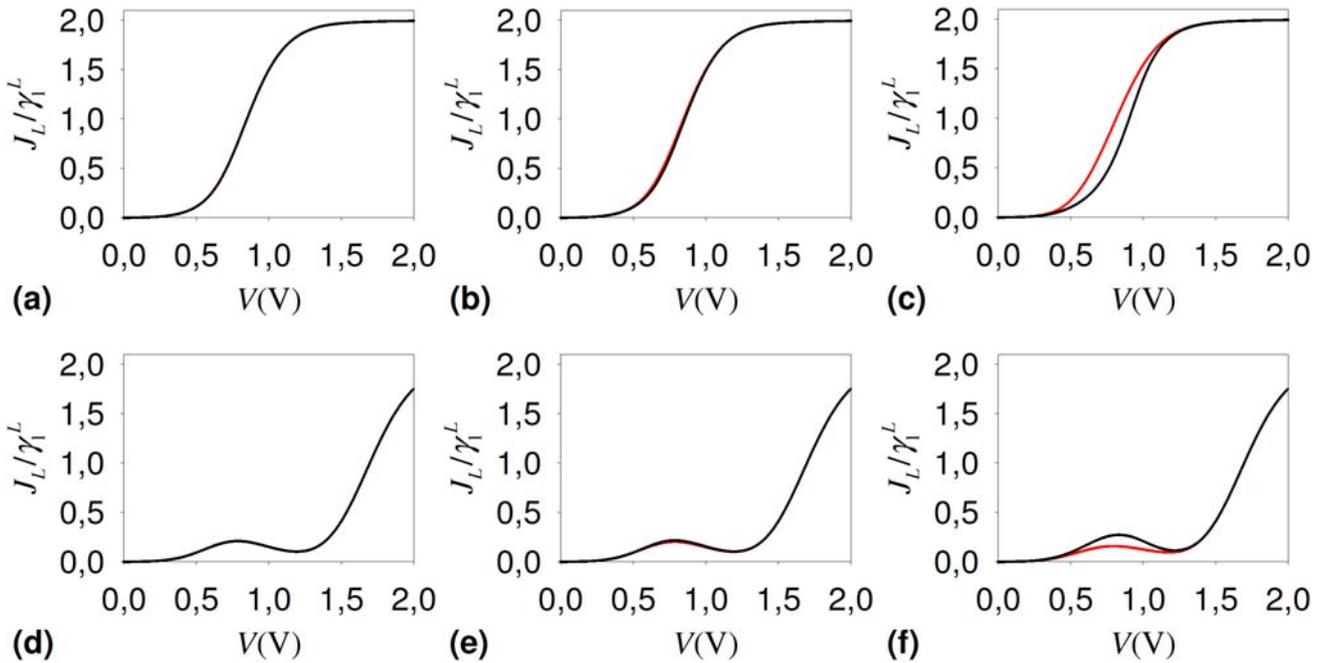

**Figure 5.** $J_L/\gamma_1^L$ plotted against $V$ over a voltage cycle, eq 23. The forward and backward sweeps are in black and red, respectively (they are on top of each other in panels a, b, d, e). The following model parameters are used: $\gamma_1^L = \gamma_1^R = \gamma_2^L = 100\gamma_2^R$, $T = 298\,\text{K}$, $E_{AB} - \mu = 0.15\,\text{eV}$, $E_{AC} - \mu = 0.3\,\text{eV}$, $\lambda_1 = 0.25\,\text{eV}$, $\lambda_2 = 0.5\,\text{eV}$, $\kappa = 4$. $\chi$ is zero in panels a-c and is 0.5 eV in panels d-f. $u/\gamma_1^L$ is $2 \cdot 10^{-5}\,\text{V}$ (left panels), $2 \cdot 10^{-4}\,\text{V}$ (center panels), and $2 \cdot 10^{-3}\,\text{V}$ (right panels).

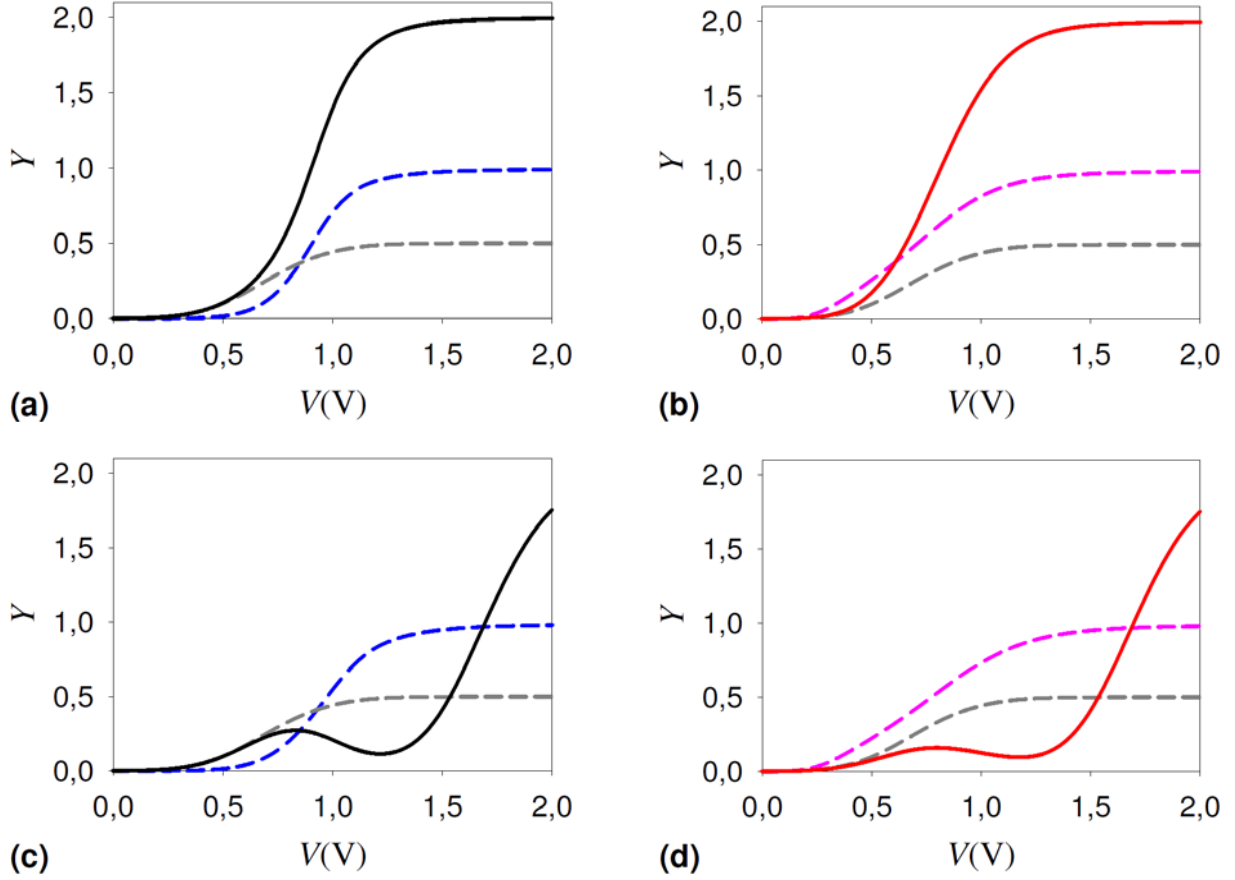

**Figure 6.** (a) The $Y$ axis represents the left-terminal current, $J_L/\gamma_1^L$ (black solid line), the steady-state current through channel 1 for empty redox site, $J_{AB}/\gamma_1^L$ (gray dashed line), and the redox site occupation $P$ (blue dashed line), plotted as functions of $V$ during a forward voltage sweep, for the same parameters as in Figure 5c. (b) Same observables during the backward sweep. $J_L/\gamma_1^L$ and $P$ are represented by red solid and pink dashed lines, respectively. (c-d) Same as a-b with the parameters of Figure 5f. Note that $J_{AB}/\gamma_1^L$ is the same in all figures.

To understand these behaviors, we consider in Figure 6 the voltage dependences of $P$ and $J_{AB}$ that appear in the main contribution $J^P$ to the current, eq 23b, for the situations of Figures 5c and 5f. Figures 6a-b focus on the case $\chi = 0$, while the case $\chi = 0.5$ eV is shown in Figures 6c-d. The following points are notable:

(a) The comparison of Figures 6a and 6b displays the hysteresis in the redox state of the molecule during a bias cycle, which results from the delay in the evolution of $P$ with respect to $P_{ss}$. In fact during the forward sweep $P$ remains negligible over a bias range wider than that predicted by eq 26 for $P_{ss}$. Similarly, during the backward sweep, $P$ takes the plateau value $P_{ss}(V \to \infty) \approx 1$ over a voltage range wider than that pertaining to $P_{ss}$. Therefore, the switch of the transport channel 1 from the conduction mode I (empty redox site and coupling strengths $\gamma_1^L = \gamma_1^R$ to the electrodes) to the more conductive mode II (occupied redox site and coupling strength $\bar{\gamma}_1^K = \kappa \gamma_1^K$ to the $K = L, R$ lead, with $\kappa = 4$) during the forward sweep occurs at higher bias voltages than the reverse switch in the backward sweep. Consequently, counterclockwise hysteresis (current is smaller in the forward voltage sweep than in the backward direction) is observed in Figure 5c.

(b) Clockwise hysteresis can be obtained if the system starts from state $C$ ($P = 1$). Moreover, if $P$ is not zero at the end of a voltage cycle (*e.g.*, see Figure S1a in the Supporting Information), the system can end a single realization of the voltage cycle and start the next one in the conduction mode II. This kind of behavior is observed, *e.g.*, in the experiments of ref 51 (see *I-V* curves in Figure 3 therein). Clockwise hysteresis loops are also found if $\kappa < 1$, namely, if the conductance of the reduced bridge is smaller than that of the uncharged molecule (examples are given in Supporting Information). This prompts future tests of our model against experimental data[4, 7, 51] that show occurrence of clockwise and/or counterclockwise hysteresis loops, as well as the possibility to predict similar behaviors in redox junctions manufactured so to fit within suitable parameter ranges.

(c) The following physical interpretation of NDR emerges from Figures 6c-d. Starting with the molecule in state $A$ and focusing for example on the current at the left interface, $J_L \cong J_{AB}$ at sufficiently low biases where $P$ is negligible, as predicted by eq 23b. As $V$ increases, $P$ becomes appreciable and consequently channel 1 can switch with probability $P$ to the conduction mode II. At higher bias voltages this switch will lead to a current $J_{\bar{A}\bar{B}} > J_{AB}$, but, because the threshold voltage[13] of mode II is higher by $2\chi = 1\mathrm{eV}$ than that of mode I, the current will decrease (NDR) before starting to rise again around the threshold bias voltage of mode II, $2(\lambda_1 + E_{AB} - \mu + \chi)/e$, and finally reaching the high-voltage plateau as $J_L \cong J_{\overline{AB}}$. During the backward sweep, because of the memory effects in the evolution of $P$, conduction mode II remains significantly populated over the voltage range in which the current is appreciable, and $J_L$ is accordingly closer to $J_{\overline{AB}}$ than in the forward sweep, with little NDR (or no NDR for sufficiently high scan rate).

(d) For zero or small enough $\chi$, $2(\lambda_1 + E_{AB} - \mu + \chi)/e$ is smaller than the threshold bias voltage for molecular charging. Thereby, conduction mode II is accessible where the redox site begins to be populated, which means that the current rises from $J_{AB}$ to $J_{\overline{AB}}$ without NDR (Figures 5a-c).

The connection between the sweep rate and the appearance of hysteresis is investigated in Figure 7. Figure 7a shows the threshold sweep rate for hysteresis, $u_0(V)$, together with the voltage sweep rates $u_1/\gamma_1^L = 2 \cdot 10^{-4}\,\text{V}$ and $u_2/\gamma_1^L = 2 \cdot 10^{-3}\,\text{V}$ used in Figures 5b and 5c, respectively. $u_1$ is smaller than $u_0(V)$ at each bias, but $u_1$ is an appreciable fraction of $u_0(V)$ over the voltage range in which the rate of electron injection $R_{AB}^L$ starts not to be negligible compared to the rate of electron delivery $R_{BA}^R$ (see Figure S2 in the Supporting Information) and thus, according to eqs 13c and 23a-b, the current is appreciable. Consequently, hysteresis appears in Figure 5b, although it is barely visible. In the same voltage range $u_2$ is larger than $u_0(V)$ so that considerable hysteresis occurs in the case of Figure 5c. Further insight into the hysteretic behavior of the molecular system is gained by the analysis in Figures 7b-c, where the appearance of hysteresis in the current-voltage response is related to the irreversible evolution of $P$, as described by eq 25 for large enough $u$ values. In these two panels we report the evolutions of molecular reduction, $dP/dt$, its reversible component, $dP_{ss}/dt = u\,dP_{ss}/dV$, and its irreversible part, $dQ/dt$, for $u = u_1$ and $u_2$. Since $dP/dt$ is of the order of $u\,dP_{ss}/dV$, the maximum of this rate in Figure 7c is about a factor $u_2/u_1 = 10$ larger than that in Figure 7b. Furthermore, the relative deviation $Q/P_{ss}$ increases considerably with the sweep rate $u$. Since the memory effects cause a delay in the evolution of $P$ compared to $P_{ss}$, the delay in charging the molecule corresponds to accumulation of the negative deviation $Q = \int dQ/dt$, whereas the delay in the achievement of full reduction (*i.e.*, the high-$V$ plateau of $P$) is responsible for decrease in $Q$. Significant memory effects occur mainly over the bias range in which $u$ is larger than or of the same order as $u_0(V)$. Nevertheless, the accumulation of memory in the response of the molecular system to the changing voltage can begin before it becomes actually observable, where the current and $P$ are both negligible but $u$ is of the order of magnitude of $u_0$, so that no hysteresis can be seen although $Q$ is an appreciable fraction of $P$ (see, for example, the low-voltage range where the sweep rate and the threshold rate $u_0$ are comparable in Figure 7a, but the current in Figure 5c and $Q$ in Figure 7c are still negligible). In addition to this, the accumulated memory affects the current-voltage characteristic over a $V$ range where $u < u_0$ but $Q$ has not been fully dissipated yet (see the high-voltage tail of the positive $Q$ peak in Figure 7c and compare with Figure 7a).

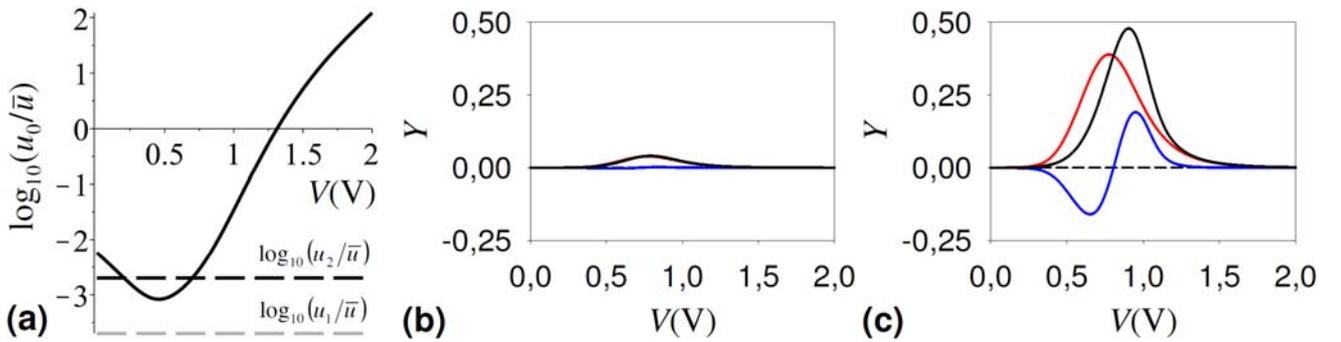

**Figure 7.** (a) The threshold voltage sweep rate, expressed as $\log_{10}(u_0/\bar{u})$ with $\bar{u}/\gamma_1^L = 1\,\mathrm{V}$ (black solid line) plotted against the voltage, using the same model parameters as in panels 6b-c. The sweep rates $u_1/\gamma_1^L = 2\cdot10^{-4}\,\mathrm{V}$ and $u_2/\gamma_1^L = 2\cdot10^{-3}\,\mathrm{V}$ are also displayed as $\log_{10}(u_1/\bar{u})$ and $\log_{10}(u_2/\bar{u})$ (horizontal gray and black dashed lines, respectively). (b-c) $Y = dP/dt$ (black), $dP_{ss}/dt$ (red), and $dQ/dt$ (blue) versus $V$ during the forward bias sweep, for $u = u_1$ and $u_2$, respectively. The time unit is $100/\gamma_1^L$. $Y = 0$ is marked by the dashed line.

The above discussion has been focused on the *average* response of the observed system over many similar bias sweeps, yet it allows direct comparison with experiments consisting in single or few voltage sweeps.[1, 4, 7, 52, 39, 51] Of particular interest is the comparison between the ensemble average and the individual sweeps. Such a comparison is shown in Figure 8 (see computational details in the Supporting Information). Figure 8a shows the average current-voltage characteristics for reversible and irreversible behaviors (as determined by the sweep rate), while Fig 8b shows the corresponding results for a single realization, where the transition probabilities $R_{AC}\Delta t$ and $R_{CA}\Delta t$, with $\Delta t$ chosen as a suitable simulation time step, are used to generate a single trajectory according to the stochastic simulation procedure[53] detailed in the Supporting Information. Of particular interest are the qualitatively different behaviors of single realizations depending on the scan rate. At fast scan rates (red and black curves in Figure 8b, whose averages over many realizations yield the red and black curves in Figure 8a), stochastic hysteresis is seen, with a single jump to the high-conductance mode in the upward run and persistence of this mode in the downward run. The single jump takes place at different biases in different sweeps, leading to the average hysteresis cycle of Figure 8a. In contrast, in the slower bias scan that on the average yields the grey curve of Figure 8a, the single realization is characterized by multiple switching between the two conductions modes of channel 1 leading to the appearance of telegraphic current noise.

The results in Figures 5 and 8 compare well with experiments such those in ref 7 (in particular, compare Figure 8b with Figure 4 in ref 7) and refs 1, 51. In ref 7 the voltage change is implemented in steps of duration $\Delta t$ referred to as current measurement integration time), so that our voltage sweep rate $u$ is proportional to $\Delta t^{-1}$. The observation[7] of hysteresis at small $\Delta t$ (0.64 ms) and telegraphic noise that averages to no hysteresis at large $\Delta t$ (320 ms) corresponds to the results shown in Figure 8.[54] Obviously, this agreement with observation does not provide a detailed description of the particular experiment, but, rather, shows the generic nature of the phenomenon and the ability to reproduce the experimental data with a generic model.

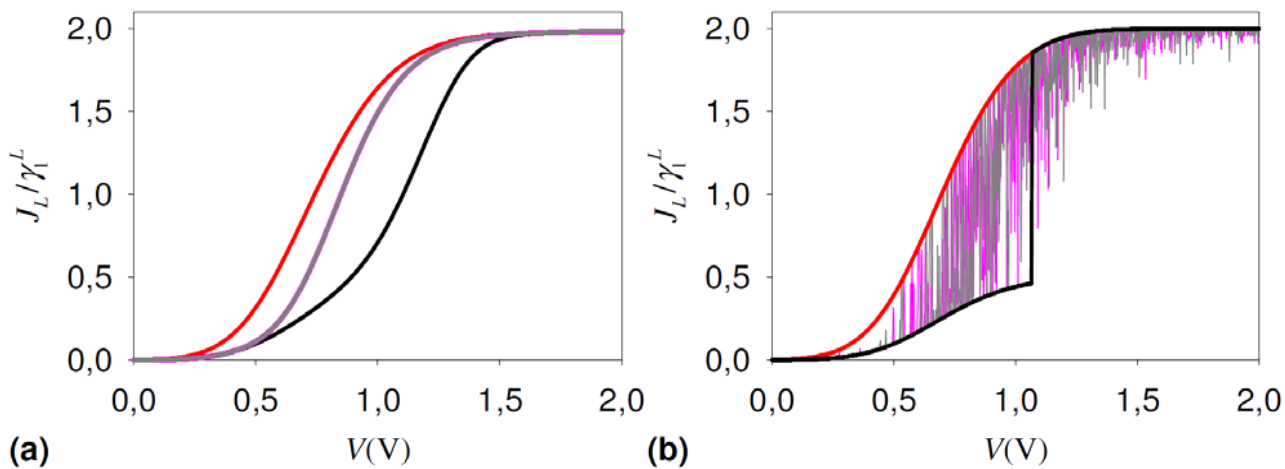

**Figure 8.** $J_L/\gamma_1^L$ plotted against $V$, using the parameters of Figure 5 with $\chi = 0$, except that $u/\gamma_1^L = 2 \cdot 10^{-2}$ V for the black and red lines while $u/\gamma_1^L = 2 \cdot 10^{-5}$ V for the grey and pink lines (forward and backward sweeps, respectively). (a) Average over many realizations. The grey and pink lines are on top of each other (reversible behavior). (b) A single realization of the system behavior over a single voltage cycle. The small direct contribution from channel 2 is neglected in this calculation.

In summary, the presence of a redox center on the molecular bridge leads to system response on a slow timescale that gives rise to hysteresis and NDR phenomena when the voltage changes on this timescale or faster. As seen in Figure 5 (see also examples reported in Figures S6-7 of the Supporting Information), the reorganization energies involved in the interfacial ET processes play an important role in determining and shaping this behavior. On the timescale considered, the system displays a bistable behavior which is enhanced by the additional stabilization provided by this reorganization. Another manifestation of this bistability is the appearance of telegraphic noise in single, slow, potential sweeps as seen in Figure 8.

**A Landauer-Büttiker-Marcus model of redox junctions.**

The two conduction channels model used in the previous section can be seen as a simplification of a quantum transport problem described by a model of two interacting transport channels (e.g. a bridge comprising two single-electron levels, each of them coupled to the leads, with Coulomb interactions between the electronic populations on these two levels) characterized by given couplings to the leads and to the phonon environment. As already noted, such models were discussed in conjunction with single electron counting measurements using point contact detectors.[14, 15, 16] In the previous section we have assumed that the molecule-lead couplings in both channels are small enough to allow treatment by classical kinetic equations and, furthermore, that electron-phonon coupling is large and temperature is high enough so that Marcus rates can be used in these kinetic equations. Here we examine another limit, where transport in the "slow" channel is assumed to be described by Marcus kinetic equations as before, however in the fast channel molecule-lead coupling is assumed to be large enough so that transport in this channel is a coherent co-tunneling process that can be described by the standard Landauer- Büttiker theory.[21, 22] As in any mixed quantum-classical dynamics, this level of description has its own intricacies and is treated in what follows with further approximations. A comparison with a numerical calculation based on the pseudo-particle Green function formalism[55] will be presented in a subsequent publication.

An approximate kinetic description of this limit can be obtained by assuming that on the timescale of interest the system can be in two states: one, denoted as state $S1$, where the slow channel 2 - the molecular redox site - is occupied, and the other where it is not (state $S0$). In terms of the probabilities of the four states in eq 19, the probabilities that the system is in states $S1$ and $S0$ are

$$P_{S1} \equiv P = P_C + P_D \qquad (28a)$$

$$P_{S0} = 1 - P = P_A + P_B \qquad (28b)$$

In each of these states, the current $I_1$ through the fast channel 1, as well as the average bridge population $<n_1>$ in this channel, are assumed to be given by the standard Landauer theory, disregarding the effect of electron-phonon interaction,[21, 22]

$$I_1(V;\varepsilon_1) = \frac{e}{\pi\hbar}\int_{-\infty}^{+\infty}d\varepsilon \frac{\Gamma_1^L \Gamma_1^R}{(\Gamma_1/2)^2 + (\varepsilon - \varepsilon_1)^2}[f_L(\varepsilon;V) - f_R(\varepsilon;V)] \qquad (29)$$

$$<n_1(V;\varepsilon_1)> = \frac{1}{2\pi}\int_{-\infty}^{+\infty}d\varepsilon \frac{f_L(\varepsilon;V)\Gamma_1^L + f_R(\varepsilon;V)\Gamma_1^R}{(\Gamma_1/2)^2 + (\varepsilon - \varepsilon_1)^2} \qquad (30)$$

where $f_K$ ($K = L, R$) denotes the Fermi-Dirac function of the $K$ electrode ($K = L, R$) and $\Gamma_1 = \Gamma_1^L + \Gamma_1^R$, and where $\varepsilon_1$ and $\Gamma_1^K$ take the values $\varepsilon_1 = \varepsilon_1^{(0)} = E_{AB}$, $\Gamma_1^K = \Gamma_1^{K0} \equiv \hbar \gamma_1^K$ in state $S0$ and $\varepsilon_1 = \varepsilon_1^{(1)} = \varepsilon_1^{(0)} + \chi$, $\Gamma_1^K = \kappa \Gamma_1^{K0}$ in state $S1$. The switching kinetics in channels 2 is described by

$$\frac{dP}{dt} = (1-P)<k_{S0 \to S1}> - P<k_{S1 \to S0}> \tag{31}$$

where for the average switching rates we invoke one of the following models:

*Model A.* The rates are written as weighted averages over the populations 0 and 1 of channel 1,[46] with respective weights $1 - <n_1>$ and $<n_1>$:

$$<k_{S0 \to S1}> = (1 - <n_1>_{S0})R_{AC} + <n_1>_{S0} R_{BD} \tag{32a}$$

$$<k_{S1 \to S0}> = (1 - <n_1>_{S1})R_{CA} + <n_1>_{S1} R_{DB} \tag{32b}$$

where $R_{AC}$, $R_{CA}$, $R_{BD}$, $R_{DB}$ are the Marcus rates defined in the above section (see discussion of eq 19 and Figure 4).

*Model B.* The rates are written as Marcus ET rates between the two system states $S0$ and $S1$, whose energy difference is taken to be $E_{S1} - E_{S0} = \left(\varepsilon_1^{(1)}<n_1>_{S1} + \varepsilon_2^{(0)}\right) - \varepsilon_1^{(0)}<n_1>_{S0} \cong \varepsilon_2^{(0)} + <n_1>_{S1} \chi$, where $\varepsilon_2^{(0)} \equiv E_{AC} \equiv E_C - E_A$.

These two models are associated with different physical pictures. Model A assumes that the switching rates see the instantaneous population in channel 1, while model B assumes that these switching rates are sensitive only to the average population $<n_1>$. Model B suffers from an additional ambiguity: the apparent change in the number of electrons on the molecular bridge between the two states $S0$ and $S1$ is $\Delta n = <n_1>_{S1} + 1 - <n_1>_{S0} \neq 1$. Nevertheless, Marcus-type rates for transferring one electron between the metal and the molecule are calculated. A discussion of these models and their validity in comparison to a quasi-exact calculation will be given elsewhere.

For both models we assume that the potential sweep is slow enough so that the above rates follow it adiabatically. As before, the time evolution of $P$, eq 31, over a voltage sweep can lead to hysteresis in the current response over a bias voltage cycle if the scan rate $u$ is fast enough. The average (over many sweep cycles with similar initial conditions) current is given by

$$I(V) \equiv eJ(V) = [1 - P(V)]I_1\left(V; \varepsilon_1^{(0)}\right) + P(V)I_1\left(V; \varepsilon_1^{(1)}\right) \quad (33)$$

where the contribution of channel 2 to the observed current is neglected. Results based on eqs 29, 31 and 33, using models A and B for the redox reaction rates, are shown in Figure 9 (see the implementation of these equations in the Supporting Information), while a single realization of the bias sweep will show telegraphic noise similar to that seen in Figure 8. The *I-V* responses predicted by both models A and B, similarly to those arising from the kinetic model in the previous section (see Figure 5), show hysteresis and hysteretic NDR. However, the considerable quantitative differences (for example, unlike in the full hopping model, no NDR occurs unless $\chi$ is sufficiently large and/or $\kappa$ is small enough) offer the possibility to discriminate between the conduction mechanisms corresponding to the two classes of models in their application to experiments. This may have relevant implications not only for the study of specific systems, but also for a more general classification of the redox molecular systems currently used in nanoelectronic experiments, based on the few global parameters characterizing the above models.

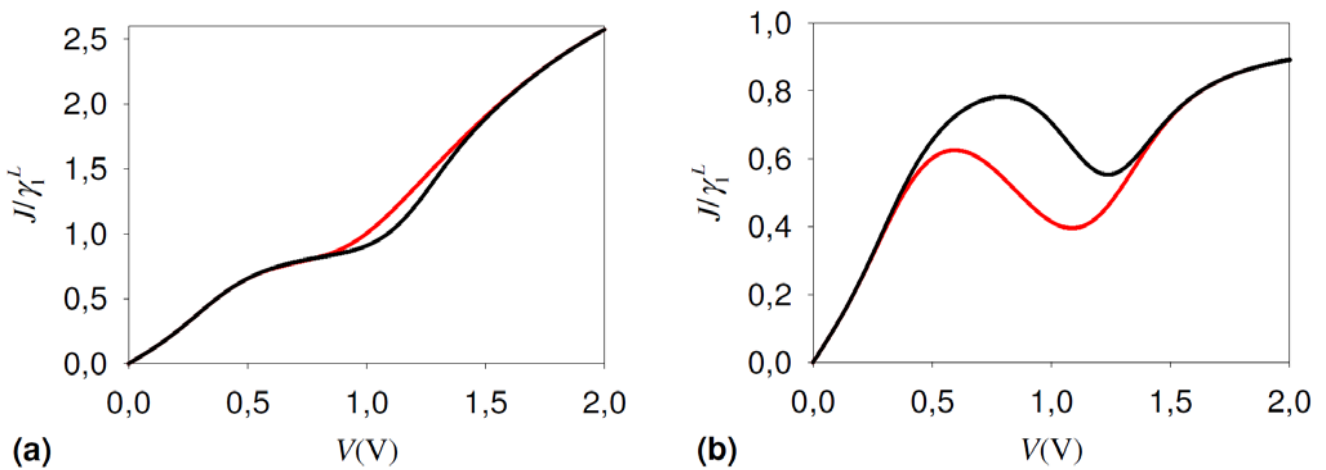

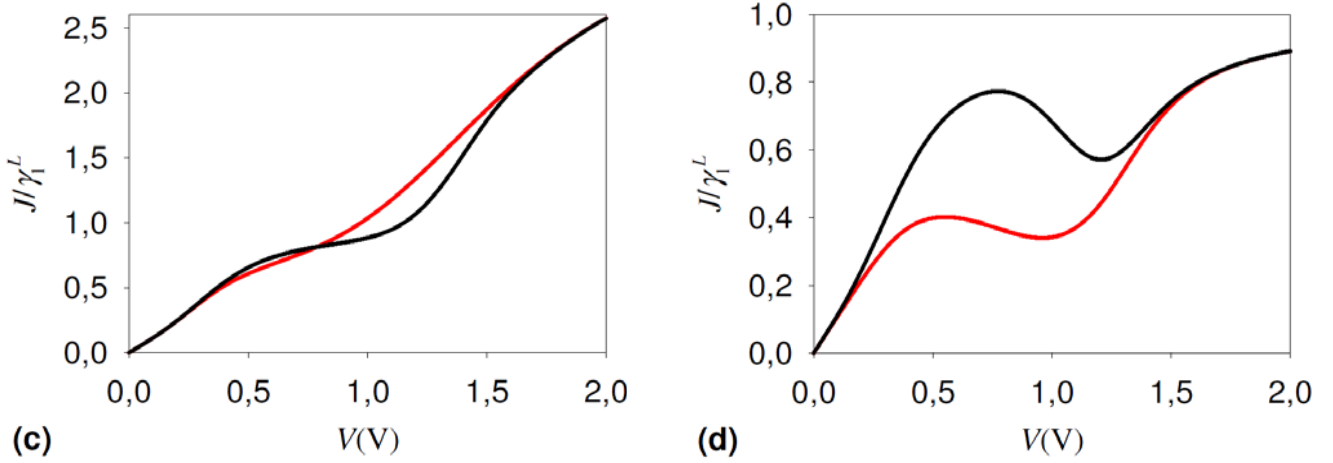

**Figure 9.** Dimensionless current $J/\gamma_1^L$ versus voltage $V$ over a bias cycle, using eq 33 with $P$ evolved according to eq 31. The forward and backward sweeps are in black and red, respectively. Models A and B for the average rates of transition between states $S0$ and $S1$ are used in a-b and c-d, respectively. The following model parameters are employed: $T = 298\,\text{K}$, $\varepsilon_1^{(0)} - \mu = 0.15\,\text{eV}$, $\varepsilon_2^{(0)} - \mu = 0.3\,\text{eV}$, $\lambda_1 = 0$, $\lambda_2 = 0.5\,\text{eV}$, $\Gamma_1^L = \Gamma_1^R = 0.1\,\text{eV}$, $\gamma_2^L = 100\gamma_2^R$, $\chi = 0.5\,\text{eV}$, $u/\gamma_2^L = \tfrac{2}{3}\cdot 10^{-2}\,\text{V}$. The dimensionless electrode coupling parameter $\kappa$ takes the value 4 in panels a and c, and 1 in panels b and d.

## CONCLUSIONS

The irreversible behavior characteristic of sufficiently fast voltammograms, expressed by distortion and shift[39] and the appearance of hysteresis and hysteretic NDR in the current-voltage response of molecular conduction junctions[9] are manifestations of the interplay between two time scales: the observation time and the characteristic time for switching between different charging states of the molecular system. The presence, on the molecular system, of redox centers on which electron localization is stabilized by a polar environment serves to separate the timescale of (slow) charging-discharging transitions from that associated with the current flow that is relatively fast even for the smallest observable currents. This localization plays a crucial role in the appearance of irreversibility effects in the range of commonly used voltage sweep rates.[6, 7, 9, 13, 39] This paper analyzes a simple generic spinless model for this phenomenon that accounts for a broad range of observed behaviors. In what follows we summarize the key features of the model and its implications:

(a) The molecular system can exhibit at least two (relatively long-lived) oxidation states characterized by charge localization in a redox site. Consequently, four distinct molecular states are considered in the kinetic version of the model.

(b) The transient localization of transferring charge and its stabilization by environmental polarization corresponds to the presence of a slow charge transport channel characterized by small interfacial ET rates. We have assumed that these rates are given by the Marcus theory of heterogeneous electron transfer,[18, 19] implying full equilibration of the environmental polarization response on the timescale of the observed kinetics and highlighting the role played by solvent reorganization about the molecular bridge.

(c) We have studied the onset of irreversibility, expressed by the appearance of bistability and hysteresis in the current/voltage response. It should be emphasized that the bistability alluded to in this paper is a transient phenomenon, characterized by the timescale of the slow channel. Its observation is determined by the voltage scan rate as compared with the rate of these charging transitions, while the observed current is determined by the second "fast" channel whose transmission properties depend on the occupation state of the redox center.

(d) Simple criteria (see eqs 10, 11, 12, 17, and 27) were obtained for the departure of the I-V response of the junction from steady-state behavior as dependent on the bias sweep rate. Accordingly, the first appearance of hysteretic behavior and the bias voltage range where it is predominantly observed can be rationalized and "predicted" from the steady-state response of the system. In principle, this may provide a route to control and modulate the junction response properties, in particular, memory effects of interest to nanoelectronics applications.

(e) The effect of solvent reorganization on the appearance of irreversibility at a single metal-molecule interface and in a molecular redox junction is analyzed and its role in the occurrence of hysteresis, NDR and hysteretic NDR is explicitly described. This can suggest suitable choices of junction components for tailoring specific features of the current-voltage response.

(f) While much of our analysis was based on a classical kinetics model with Marcus rates, quantum

coherent transport in the fast channel has been considered as well. Such model shows qualitatively similar behavior, with significant quantitative differences. It also raises important issues in the approximate description of the mixed quantum classical dynamics that will be further discussed in a future publication.

Finally, we wish to note that the simplicity and generic character of the presented models may provide a useful framework for further theoretical developments, including the consideration of situations where neither the hopping nor the fully coherent mechanisms are appropriate to describe the conduction via the effective transport channels.

**ACKNOWLEDGEMENTS.** This research was supported by the Israel Science Foundation, the Israel-US Binational Science Foundation and the European Research Council under the European Union's Seventh Framework Program (FP7/2007-2013; ERC grant agreement n° 226628). We wish to thank Michael Galperin for fruitful discussions.

**APPENDIX**

**Marcus ET rates and their approximation near** $V = (E_{AB} - \mu)/e$. The interfacial ET rates can be written as the following sums of analytic functions:[13, 30]

$$R_{AB} = \frac{\gamma}{4} S(\lambda, T, \alpha) \exp\left[-\frac{(\alpha - \lambda)^2}{4\lambda k_B T}\right], \qquad R_{BA} = \frac{\gamma}{4} S(\lambda, T, \alpha) \exp\left[-\frac{(\alpha + \lambda)^2}{4\lambda k_B T}\right], \qquad (34a)$$

where

$$\alpha \equiv \mu - E_{AB} + eV, \qquad (34b)$$

$$S(\lambda, T, \alpha) = \sum_{n=0}^{\mathcal{N}} \frac{1}{2^n} \sum_{j=0}^{n} (-1)^j \binom{n}{j} \left[\eta_j(\lambda, T, \alpha) + \eta_j(\lambda, T, -\alpha)\right], \qquad (34c)$$

$$\eta_j(\lambda, T, \alpha) = \exp\left\{\frac{[(2j+1)\lambda + \alpha]^2}{4\lambda k_B T}\right\} \operatorname{erfc}\left[\frac{(2j+1)\lambda + \alpha}{2\sqrt{\lambda k_B T}}\right], \qquad (34d)$$

and the limit superior $\mathcal{N}$ truncates the otherwise infinite sums. In eq 34, $\gamma$ is the coupling strength to the metal, taken as a constant. $k_B$ is the Boltzamann constant, $T$ is the temperature, $\lambda$ is the reorganization energy of the molecular system (including the solvent), and $E_{AB} \equiv E_B - E_A$, where $E_A$ and $E_B$ are the energies of the states $A$ and $B$, respectively.

For $|eV - E_{AB} + \mu| \ll \lambda$ and $\lambda \gg k_B T$, one can write the above ET rates in the Gaussian-like form proposed my Marcus.[56] Therefore, it is

$$R_{AB}\left(\frac{E_{AB} - \mu}{e}\right) = R_{BA}\left(\frac{E_{AB} - \mu}{e}\right) = \frac{\gamma}{4} S(\lambda, T, 0) \exp\left(-\frac{\lambda}{4 k_B T}\right). \tag{35}$$

$S(\lambda, T, 0)$ is truncated as in eq 34c for any practical calculation. However, in the exact limit $\mathcal{N} \to \infty$, it can also be recast as[30, 57]

$$S(\lambda, T, 0) = 4 \sum_{j=0}^{\infty} (-1)^j \eta_j(\lambda, T, 0) = 4 \sum_{j=0}^{\infty} (-1)^j \exp\left[\frac{(2j+1)^2 \lambda}{4 k_B T}\right] \mathrm{erfc}\left[\frac{2j+1}{2} \sqrt{\frac{\lambda}{k_B T}}\right]. \tag{36}$$

For $\lambda \gg k_B T$, one can use the asymptotic expansion of the complementary error function and write

$$\begin{aligned}
S(\lambda, T, 0) &\cong 4 \sum_{j=0}^{\infty} (-1)^j \exp\left[\frac{(2j+1)^2 \lambda}{4 k_B T}\right] \frac{2}{2j+1} \sqrt{\frac{k_B T}{\pi \lambda}} \exp\left[-\frac{(2j+1)^2 \lambda}{4 k_B T}\right] \\
&= 8 \sqrt{\frac{k_B T}{\pi \lambda}} \sum_{j=0}^{\infty} (-1)^j \frac{1}{2j+1} = 8 \sqrt{\frac{k_B T}{\pi \lambda}} \sum_{j=1}^{\infty} (-1)^{j+1} \frac{1}{2j-1}.
\end{aligned} \tag{37}$$

Then, using the equation (where $E_{2n}$ are Euler numbers)[58]

$$\sum_{j=1}^{\infty} (-1)^{j+1} \frac{1}{(2j-1)^{2n+1}} = \frac{\pi^{2n+1}}{2^{2n+2} (2n)!} |E_{2n}| \tag{38}$$

with $n = 0$ (hence, $E_0 = 1$), we obtain

$$S(\lambda, T, 0) \cong 8 \sqrt{\frac{k_B T}{\pi \lambda}} \sum_{j=1}^{\infty} (-1)^{j+1} \frac{1}{2j-1} = 2 \sqrt{\frac{\pi k_B T}{\lambda}}, \tag{39}$$

hence the expression of $\rho$ for $\lambda \neq 0$ in eq 9 after substitution in eq 35 and use of eq 3b. This analysis can clearly be applied to each interface of a junction, with $V$ replaced by $|\phi_K|$ ($K = L$ or $R$).

## ASSOCIATED CONTENT

## Supporting Information

Insights into the connection between voltage sweep rate ad reversible/irreversible behavior; typical lead-molecule coupling strengths deduced from experiments; analytical expressions of the threshold scan rate for appearance of irreversibility; additional applications of the two and four state redox junction models; generalizations of the four-state hopping model with Marcus rates; implementation of four-state Marcus model and Landauer-Büttiker-Marcus model. This material is available free of charge via the internet at http://pubs.acs.org.